# On the usability of generative AI: Human generative AI

Anna Ravera[1], Cristina Gena[2,†]

[1]*School of ICT and Media, University of Turin, Italy*
[2]*Dept. of Computer Science, University of Turin, Italy*

**Abstract**
Generative AI systems are transforming content creation, but their usability remains a key challenge. This paper examines usability factors such as user experience, transparency, control, and cognitive load. Common challenges include unpredictability and difficulties in fine-tuning outputs. We review evaluation metrics like efficiency, learnability, and satisfaction, highlighting best practices from various domains. Improving interpretability, intuitive interfaces, and user feedback can enhance usability, making generative AI more accessible and effective.

## 1. Introduction and background

Today's generative AI systems are increasingly widespread and despite continuous updates and new features they present significant usability issues. Among the primary problems observed in the analysis of the most generative AI systems, control during interaction stands out [1]. The issue of user control versus automation focuses on AI and is crucial in every human-computer interaction design. It applies to a vast range of objects and processes involving some degree of automation, such as household appliances, smartphones, and car safety devices. It is essential to design the experience so that the level of control users have over operations is clear. Each case is unique, and it is not always true that greater user or machine control is the best solution.

Specifically, in AI, the analysis of control and interaction involves the dual role of AI systems as both "assistive tools" and "colleagues" [2, 3, 4]. This leads to the ideal scenario of "Human-Machine Teaming," where humans and AI collaborate to achieve a common goal. The balance is achieved through the Hybrid Intelligence approach, which surpasses the CITL and HITL paradigms (Computer in the Loop and Human in the Loop). This approach enables the accomplishment of complex goals by combining human and artificial intelligence to collectively achieve superior results, improving continuously through shared learning [5]. Some significant usability issues stem from the paradigm shift in human-AI interaction and the failure of interfaces to adapt to and to personalize to new user needs [6, 7, 8, 9]. Interaction has shifted from a GUI-based model within the broader command-based interaction paradigm—where users reach the desired result independently through commands and hypertext links [10] to an intent-based model [11].

In command-based interaction, users do not necessarily need a clear idea of what they want to achieve at the start; they can refine their search and define the desired content progressively, as the state evolves after each command [12]. In contrast, interaction with AI systems involves a paradigm where users express the desired outcome without specifying "how" to achieve it. This reverses the locus of control in the interaction.

Interaction can still be iterative when the AI's output does not meet user expectations, allowing users to refine or modify the result. This is an intent-based interaction. However, difficulties in obtaining the desired outcome arise from both the AI's interpretation and the translation of intentions into prompts. An evolution in the user experience of AI systems is necessary, integrating GUI-like characteristics with intent-based interaction [13].

According to [14] successful UX framework for AI systems can be composed of the following elements: Context, Interaction, and Trust. Implementing features respecting this framework can significantly enhance user experience in terms of research and interface design. Among these, trust requires particular attention. The trust relationship between humans and AI determines the success of the interaction, beyond the technical efficiency of the generated output. A successful interaction involves completing tasks without unexpected issues, errors, or unnecessary additional activities. Trust becomes the key to system adoption: if users trust the system, they will continue to use it; otherwise, they will abandon it.
Regaining lost trust is a long and difficult process.

The intent-based interaction model [15] typically involves either voice interaction or prompt-based interaction. While voice interaction improves accessibility, it is not suitable for complex tasks. Promptbased interaction, on the other hand, faces its own challenges. Despite being established as the primary mode of interaction, prompt-based interaction often fails to help users achieve satisfactory results independently without the right tools and guidance. Writing an effective prompt requires certain skills and considerations to achieve the desired outcome. The written prompt represents only the "tip of the iceberg" with the submerged portion consisting of additional information in the user's mind (or yet to be formulated). These implicit elements often play a more critical role in enabling the AI to deliver precise and in-depth content aligned with user intentions.

Another issue tied to usability relates to the black-box nature of AI systems. These systems, often generative but not exclusively, are opaque to users. Users are often unaware of how the AI arrives at its outputs, how accurate the results are, why certain content is included while others are not, and which sources were used. Unlike non-AI systems, where outputs are predetermined and based on algorithms and rules, AI systems produce outputs that may be unexpected and non-deterministic [16]. This unpredictability and lack of transparency can lead to damage to user expectations.

In this rapidly evolving AI landscape, improving system usability and user experience focuses on the conscious integration of an HCAI (Human-Centered AI) approach in interface design [1]. This approach should consider all stakeholders involved in AI systems, account for real user needs, their ability to express themselves effectively through prompts, and the effort required to adapt to new intent-based interaction systems.

This paper has been organized as follows: in Section 2 we present the shift from From Conduit Metaphor to Prompt Engineering; in Section 3 we present and discuss a prototype we design for usable generative AI systems; in Section 4 we present a usability test performed on the prototype, while in Section 5 we present the results and we conclude the paper.

## 2. From Conduit Metaphor to Prompt Engineering

Communication between humans occurs in a multimodal manner and an important part of communication as a whole is non-verbal communication, which includes body language, eye language, touch and other forms of expression without words [17]. Although these are the characteristics of human-tohuman communication, this is not the model used for the design of most technological interfaces that require machine-to-human communication.

For this type of design, one of the most famous communication models is used, which is one of the most well-known in communication theory. This model suggests that a speaker encodes their thoughts by placing them into symbols—words—and a receiver decodes these received words, assigning them meaning. This communication paradigm emerged in the 1950s, focuses on the message, and is called the conduit metaphor [18]. According to this model, meaning exists within the message itself, and communication is successful if the sender and the receiver share the same rules for encoding and decoding messages. This was one of the first models presented to describe the communication process. Today, we know that the exchange of information does not occur in this way [19], but despite this, the model has shaped the way interaction between humans and machines has often been designed, particularly between humans and linguistic interfaces, such as LLMs, and consequently, generative AI systems.

The metaphor of the conduit and the functioning of the *prompt* as a mode of interaction between AI and humans have one thing in common. In fact, in both cases it is taken for granted that the person who is communicating is clear in mind the concept he or she wants to communicate, but the reality is not so clear and defined. Indeed, in human conversation, there is support for the interlocutor to shape the message, its content and meaning continuously and it is evident that in the case of interaction with a generative AI, for example, this cannot be possible in the current state of development of technology. Indeed, during a conversation, implicit feedback is sought from the interlocutor, which can be expressed not only through words, but also through the wide range of non-verbal communication, which is just as important as verbal communication. The 'effort' of communication is distributed among the interlocutors and the responsibility for the success of the conversation is shared [20].

Today, interaction via *prompts* does not allow this type of interaction. In fact, just as interaction between human beings takes place in the manner just described, so a non-expert user approaches the use of generative AI with this innate mode of conversation. What is sought is a collaborative interaction, a shared work in which the AI behaves, precisely, like a peer with whom the user is having a conversation. There is a need for a shift towards shared control and equal contribution to the conversation, [20, 11]. These needs fully reflect the principles and goals of the *Human Centered AI* approach which allows an ethical and usable approach, especially in the long run, for a fruitful and effective human-AI interaction. In practice, this approach [20] translates into a necessary redesign of conversational interfaces taking into account HCAI principles and the nature of human communication, which is multimodal.

## 3. Interface Development

It became clear how necessary a *human-centered* approach is for the design of AI systems and beyond [1]. Then, the criticalities of the main generative AI systems present today were exposed together with an analysis of their functions and use. Next, one of the main ways of interacting with AI, the *prompt*, was presented, and *prompt engineering* techniques were analysed, all accompanied by an analysis of the ease of use of the *prompt* tool and the difficulties non-expert users encounter when interacting with AI.

At this stage, it is necessary to shift the focus of observation and focus on another of the components of the interaction that can play the role of modifying the satisfaction and success of the interaction itself and more generally improve the human-AI systems relationship: the interface.

In fact, an interface of a hypothetical generative AI system for text and image generation will be proposed below, which attempts to encapsulate the issues discussed so far, considering the *humancentered* approach, use by non-expert users, intuitiveness, collaboration and the communication mode through which humans interact.

### 3.1. The interface as a whole

With regard to the *desktop* interface as a whole, we chose to maintain a structure similar to that already present in other generative AI systems and typical of conversational interfaces even before the integration of AI. This makes it possible to maintain consistency between known interfaces of similar systems and to ensure that the user is not disoriented and can recognize buttons and functions present. The name chosen for the application is manifest of its intended purpose: H-gAI, *Human generative AI*.

The interface, see Fig. 1, presents a dialogue screen with most of the space dedicated to *chat*. On the right-hand side, at the top, there is access to the profile and profile settings and on the left-hand side the history of past *chats*. The *floating bar* on the left side can be hidden for more *chat* space. There is also a button to create a new conversation, in addition to the history and the possibility to search for a *chat* in the archive.

**Checking the system status**. During the conversation, the user is always in control of the state the system is in, an important principle of *human-centered* AI system design. In fact, from figure 1 it can be seen that there are two indicators useful to the user. The first is located inside the *box* in which the generated content is inserted and consists of the warning that the system is generating the response. The other indicator is located at the bottom of the *box* containing the generated *output* and indicates the progress status of the content, so the user is always in control of the length of the response and the level of completion of the content.

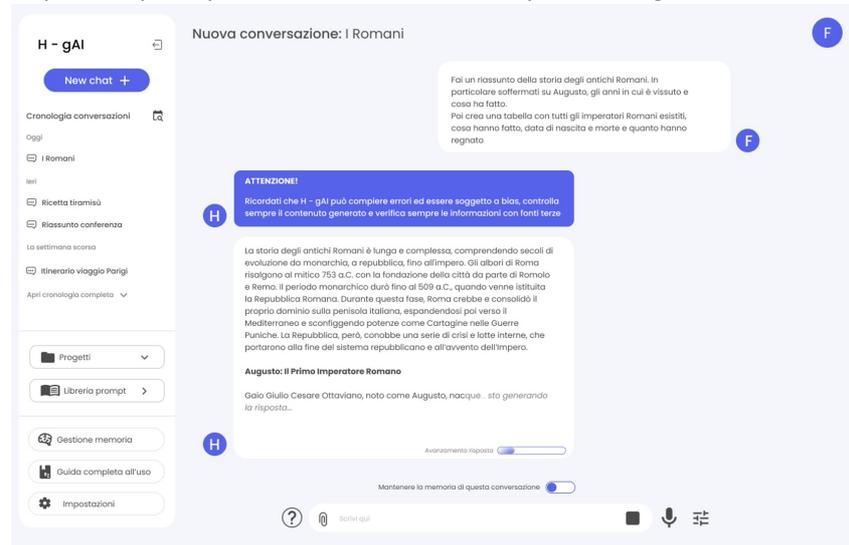

**Figure 1:** The conversation screen with the example of chat

**Warning: About errors and biased content.** A further difference from other generative AI systems is found in the first response message of the AI after the user's first *prompt* and can be seen in 1 in the purple *box* with white text. This is a *warning* message that appears every time the user starts a new conversation and warns of the possibility of the AI making mistakes and generating results that are subject to different types of *bias*. As present in the *guidelines* for the design of *human - centered* AI systems [21] it is very important to make the user aware of what the AI system can do and its limitations. In other generative AI systems analyzed the *reminder* about the possibility of making mistakes is usually found below the *input* box of the *prompt*, while we decided in this case to make it more prominent and actively propose it to the user at every interaction to remind them of the importance of *fact-checking* and not to assume as true and correct all AI generated content. At the same time it is a message that does not hinder or slow down the interaction.

Returning to the *floating sidebar*, in the portion below the history there are links to the project folder and the *prompt* library. we decided to introduce these two elements to improve workflow and user *chat* management.

**Projects**. The project folder provides access to a real archive, where the user can create folders in which to place conversations relating to the same topic, as if it were the computer's local *file* management. In the generative AI systems analyzed so far, conversations cannot be organized into themes and folders, but we think that for a continuous and wide use of AI systems it can be a useful function to keep an order in the conversations and the materials generated.

**Prompt library**. Access to the *prompt* library, on the other hand, consists of a link to a new page where there is an archive of *prompts* used and loaded by the users themselves divided into categories, so that the user using the AI system can draw inspiration from or use *prompts* already used that have generated a satisfactory result for other users previously. The last elements are found under access to projects and the *prompt* library: direct access to memory management, user's guide, settings.

**Conversational memory**. Like some of the best-known generative AI systems present today, we have introduced conversational memory. This function allows for a deeper relationship of the user with the AI system and avoids the repetition of ancillary contextual explanations that would slow down the interaction. With this function on the interaction screen the user can always check what has been stored by the AI and modify the elements in the memory at will. Still on the subject of memory and the user's full control over the items to be stored, there is a button on the conversation screen, above the *prompt* entry box, which is always present from the moment a *chat* is started, allowing the user to decide whether to keep that specific conversation in memory or not. By default, the memory is not activated, so it is up to the user to actively decide as he or she wishes.

**User's Guide**. The user's guide to the use of the AI system is essential for proper interaction and for the user to get the best out of the AI system while maintaining his or her sense of *mastery*. A comprehensive guide to all the functions present with examples and demonstrations is presented to the user the first time they register. Thereafter, it remains accessible at any time in the *floating sidebar*.

**Information and research.** To the left of the *input* box for the *prompt* is a clickable icon of a question mark, see Fig. 1. This icon gives the user access to a *box* that overlaps *the chat* and allows searching for any item, whether it be a *prompt* setting or a specific function. From this section, there is also a link to some useful features such as reporting a problem, keyboard shortcuts for some quick functions, *privacy* policy and the app's terms of service.

### 3.2. Modes of interaction

Starting from this basis, which at the same time retains some elements already present in other generative
AI interfaces and adds new ones, the reasoning for implementation proceeds both by considering the *guidelines for* the design of AI systems and the principles for the development of *human-centred* generative AI.

One of the first elements to consider concerns the mode of interaction. We kept the *prompt* interaction mode alongside the voice interaction mode.

### 3.2.1. Voice interaction

Although voice mode does not meet all the needs of use of generative AI systems, it is useful for noncomplex tasks and, as seen above, broadens the possibility of using the AI system, making interaction more accessible and intuitive even for users who have difficulty in writing. Although not present in all generative AI systems in the *desktop* version, we have chosen to include it, allowing for additional voluntary customisation by the user.

The voice interaction mode is activated by clicking on the microphone icon next to the text *input* box and allows an instant *speech-to-text* transcription of what the user says. In addition, the response provided by the AI system is also kept in writing in addition to being spoken aloud. Speech interaction preferences can be managed by *double-clicking* on the microphone icon 2.

A control panel will then open from which certain settings can be managed. It is in fact possible to select from four different types of voices, depending on the user's preference, and it is also possible to change the speed at which the voice of the AI system is played. As a final control panel setting, we have included an 'automatic voice assistance' mode.

**Voice assistance**. The voice assistance mode, which can be activated from the voice interaction control panel, allows voice assistance to intervene in the event of user difficulties.

The user can activate this function from the control panel and, during a conversation, if the AI system recognizes after sending a few *prompts* that the user is unable to express the desired content or is unable to create an effective *prompt*, the AI system will intervene by asking the user if he/she prefers to use voice interaction to better express what he/she wants to achieve. The microphone icon will then activate allowing the user to intervene. This is not, however, a function that stops the interaction; the user can disregard the AI's advice and continue writing the *prompt*.

Each control panel setting has a detailed explanation of operation that can be reached by clicking the information icon at the end of each function description.

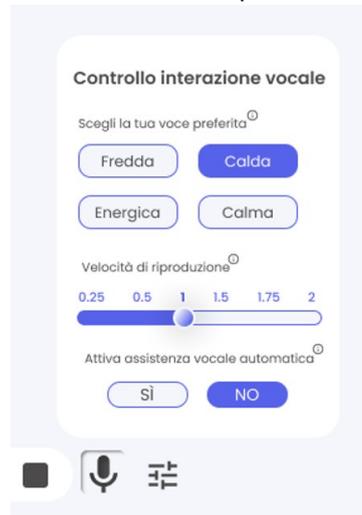

**Figure 2:** Double-click on the microphone icon to access the control panel for voice interaction

### 3.2.2. Interaction vAI *prompts*

In *designing* an AI system that would be more *human - centered* than the existing ones, we still chose to retain the interaction mode via *prompts*. The main reason for this choice is the consistency for the user and the familiarity he or she possesses with both *chatbot* interaction systems that do not integrate AI, but especially with the generative AI systems already present and in use. Moreover, it would not make sense to disrupt an interaction paradigm that has established itself as the main one in human-AI interaction.

Interaction via *prompts* occurs as in any other generative AI system: there is, in fact, a text input bar in which the user can write his/her requests and then press the enter button to generate the required content. Up to this point, nothing different from the familiar interfaces is noticeable. To improve the interaction with the AI and amplify the user's sense of control over the operation, there's a control panel for this mode of interaction as well, which is activated by pressing the icon next to the microphone. The control panel opens in a vertical section at the side of the *chat*, so as not to get in the way of the user's flow of conversation, and presents a real *dashboard* from which the user can control certain parameters of the interaction (Figure 3).

The use of the control panel is absolutely optional and the interaction can be successful even without the user controlling the parameters. The presence of this possibility helps, at the same time, to maintain its sense of *mastery* and to refine the required result.

We have divided the user-accessible controls in the control panel into three different sections: i) Basic output controls; ii) Response models; iii) Assistance in writing the prompt. Some of the reported controls, even according to what happens in generative AI systems, seem designed for experienced users. In reality, they are all first described in detail in the user guide, their use is entirely optional, but they represent useful parameters on which to intervene in order to maximize the result the user desires. **Basic *output* controls**. In the first section of the control panel, among the basic controls, we decided to include the possibility of changing the language of the generated content by opening the drop-down menu. There are some controls which can also be added directly into the *prompt* by the user, but which can be selected from a pre-selected menu. It is indeed possible to act on the style, tone and length of the content. Adjusting these parameters in advance allows for quick interaction and saves the user a lot of editing once the content has been generated. The user can find eight different types of tone, from which he/she can choose: formal, informal, persuasive, neutral (*default* option), empathetic, inspirational, didactic, humorous, and eight types of style: descriptive, formal, informal, friendly, motivational, neutral, creative, instructive.

A further function, which can be activated by the user in this section, concerns the possibility

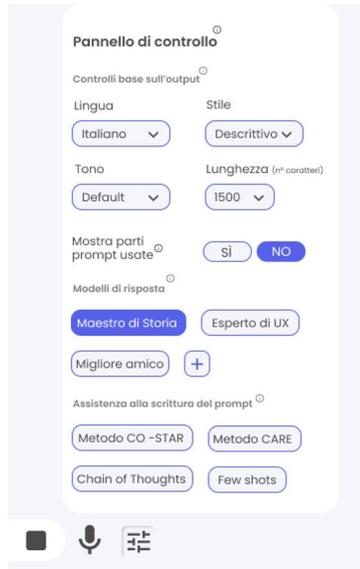

**Figure 3:** The control panel that is activated at the side of the chat and allows the user to intervene on certain parameters of the generated response

of showing, in the result generated by the AI, the parts of the *prompt* used to obtain the result to allow the user to see which parts of the input were more or less useful in the generation. This function is particularly useful in a learning phase and when approaching the generative AI system and experimenting *with prompting* techniques. Providing feedback of this kind makes it possible to improve the interaction and to understand which elements of the prompt were 'successful' in the realization of the content and which details, on the other hand, turned out to be superfluous.

**Response patterns**. As in other generative AI systems when writing the *prompt,* it is possible to request the AI to take on a certain role (*roleplay mode* ) when generating the response. We find the possibility for the user to create roles that he or she uses most often and save them in the control panel for easier and faster interaction if certain roles are used frequently.

**Prompt writing assistance**. As the last section of the control panel, the user can receive assistance when writing the *prompt*. Therefore, one of the user's difficulties in writing the *prompt* lies in remembering the important elements that must be entered for the AI system to generate an effective result and to avoid an unnecessarily long conversation. We can notice four of the simplest but comprehensive methods for producing effective *prompts* from which the user can choose. The AI system will assist him/her in composition by following him/her step by step: the two *prompt* writing *frameworks*, COSTAR [22] and CARE[1], together with two basic *prompt engineering* techniques: *Chain of Thoughts* [23] and *Few Shots* [24] (Fig. 3).

---
[1] https://www.nngroup.com/articles/careful-prompts/?lm=intelligent-assistant-usabilitypt=article

Once the AI system has selected the *prompting* method for which to receive assistance, it will provide a brief guide for the user directly in the *chat*, explaining the operation of the method as a whole and then guiding him/her through the various steps required. This type of assistance puts the user in a position to choose how to handle the *prompt*, allowing them to learn new strategies and thus optimize their interactions.

For each type of *prompt* proposed in this section, there is also a small user guide, which is activated by *hovering* over the button relating to the selected technique and offers the user a brief overview of the *prompting* technique.

### 3.2.3. End of interaction

Once the interaction between the user and the AI system is complete, it is possible to perform actions on the generated response through the clickable icons that appear at the end of the interaction below

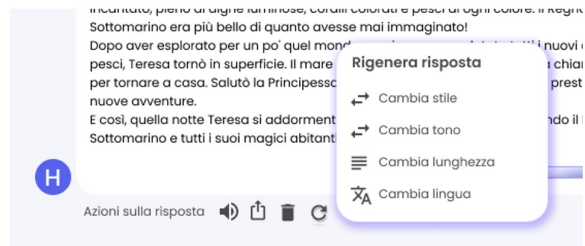

**Figure 4:** Possible actions on the generated response

the content (Fig. 4).

It is indeed possible to have a read aloud of the generated answer with a *click* on the microphone icon, it is possible to share the conversation and copy the text, delete the answer or re-generate it. In addition to the simple re-generation of the answer, it is possible to perform more precise actions with a double *click of* the pointer that opens an interaction *box.* It is indeed possible, for the response obtained, to change the style, tone, length and language with a link to the parameters in the control panel seen above.

## 4. Usability Testing

In order to test the Figma-based prototype and its functionalities and to reason about the necessary modifications, we carried out a classical usability *test* with 5 users. During the test, the users had to achieve certain goals and during the execution they had to explain aloud the actions they were performing using the *Thinking Aloud* technique. A tester was always present during the test. The test was divided into three different parts:

1. In the first part of the test, the tester introduced the participants to the application's functionality, provided a brief description, and explained the purpose of the test and the tasks they were supposed to complete. The tester also asked if they had ever used a generative AI system and addressed any questions or doubts they had. This phase lasted approximately 10 minutes.
2. During the second phase, each user performed the tasks described below. All users performed the same tasks. During the execution each user described step by step the actions they were performing using *Thinking Aloud* protocol and at the same time we took notes on their actions, expressions and movements in the interface. This phase lasted a total of 15 minutes.

3. During the third and final phase of the test, the tester asked all users for general feedback on their thoughts about the interface and how they used it; the duration of this phase was about 5 minutes.

*The tasks* assigned to users were as follows:

**Task 1**: Change the response style of the AI system: You are on the main page of the AI system and are about to start a conversation. In this *app* you can set a priori parameters for the response, we ask you to change the style of the response.

**Task 2**: Change the playback speed of reading aloud: in this AI system you can also interact with voice interaction and you can change parameters on how the AI responds to you, change the playback speed. **Task 3**: Response regeneration: you have a *chat* but you are not satisfied with the response and want to regenerate it and change the style.

**Task 4**: Search for how to delete *chat* history. You are on the AI system and want to do a general search, e.g. you want to search how to delete the history.

Upon completion of the *tasks*, we assigned each participant a score on a scale of 1 to 3, where 1 means that the task was not completed, 2 that it was completed with some difficulty, 3 that it was completed quickly and without difficulty, see Table 4. We also included the age group and an 'exp' column indicating experience, i.e. whether the users had ever used a generative AI system

| Users  | Age   | Exp | Task 1 | Task 2 | Task 3 | Task 4 |
|--------|-------|-----|--------|--------|--------|--------|
| User 1 | 18-24 | Yes | 2      | 2      | 3      | 1      |
| User 2 | 18-24 | Yes | 3      | 1      | 2      | 1      |
| User 3 | <18   | Yes | 2      | 1      | 1      | 1      |
| User 4 | 18-24 | Yes | 1      | 2      | 1      | 1      |
| User 5 | > 55  | No  | 2      | 1      | 3      | 2      |

**Table 1**
Summary table on age range of test participants, previous experience with generative AI systems and completion of assigned tasks. 1: Did not achieve the goal; 2: Achieved the goal with difficulty; 3: Achieved the goal easily

## 5. Discussion and Conclusion

The user test was very useful, providing insights into both the strengths and areas for improvement in the design. The similarity between the presented system and familiar generative AI systems or chatbots helped users recognize icons and functions easily. However, some features placed on the main screen to simplify the user experience caused confusion, as many users resorted to the settings to adjust parameters. This highlighted the necessity of having a clear and complete user guide readily available to help users understand how to perform various actions.

Task execution times were longer in the first task, averaging 52 seconds, likely due to a lack of familiarity with the interface. However, the times improved with each subsequent task, dropping to 30.2 seconds by the fourth task, indicating increasing comfort with the system. Task 4, being more complex, had the highest failure rate at 80% underscoring the importance of a more intuitive interface and clearer guidance, especially regarding the functions near the prompt input box.

The test emphasized the importance of balancing additional functions on the main screen while maintaining familiar reference points for users. This balance is crucial for user confidence and ease of navigation. The results suggest that human-centered interfaces for generative AI systems will become more common, with a focus on providing intuitive, user-friendly designs and comprehensive guides to help users fully understand the system's capabilities.

The proposal aims to apply human-centered design principles to create a feasible, user-friendly interface for generative AI systems, without disrupting existing design practices. It focuses on enhancing user control and awareness during AI interactions, highlighting both the user's ability to master the system and the potential mistakes AI can make. The goal is to foster collaboration, co-creativity, and sustainable interaction, emphasizing human-AI collaboration in a work environment. A human-centered approach will be crucial for the success of the third wave of AI, which seeks to create a holistic and ethical paradigm for human-AI collaboration, leading to better outcomes through collective efforts.